\documentstyle[12pt]{article}
\makeatletter
\newcommand{\singlespacing}{\let\CS=\@currsize\renewcommand{\baselinestretch}{1.0}\tiny\CS}
\newcommand{\doublespacing}{\let\CS=\@currsize\renewcommand{\baselinestretch}{1.5}\tiny\CS}

\begin{document}

\title{WKB and MAF quantisation rules for spatially confined quantum mechanical
systems}
\author{Anjana Sinha\thanks{e-mail:res9523 @ www.isical.ac.in} \\
and \\
Rajkumar Roychoudhury\thanks{e-mail:raj @ www.isical.ac.in} \\
{\it Physics \& Applied Mathematics Unit}\\ 
{\it Indian Statistical Institute} \\ {\it Calcutta - 700035}\\ {\it India}}

\date{}

\maketitle

\vspace*{1cm}

\centerline{\bf Abstract}

\vspace{0.3cm}

\thispagestyle{empty}

\setlength{\baselineskip}{19.5pt}

A formalism is developed to obtain the energy eigenvalues of spatially
confined quantum mechanical systems in the framework of the usual
Wentzel-Kramers-Brillouin (WKB) and Modified Airy Function (MAF) methods.
To illustrate the working rule, the techniques are applied to 3 
different cases, {\it viz.} the confined $1$-dimensional harmonic 
and quartic oscillators, and a boxed-in charged particle subject to 
an external electric field. The energies thus obtained are compared 
with those from shifted $1/N$ expansion, variational
and other methods, as well as the available exact numerical results. 

\newpage

\section*{I \ Introduction}

Spatially confined quantum mechanical systems [1-3] make an interesting study due to
their importance in a variety of physical problems - e.g. studying
thermodynamic properties of non-ideal gases, investigation of
anharmonic effects in solids, in atoms and in molecules under high pressure, 
impurity binding energies in quantum wells and near-surface donor states, and
even in the context of partially ionised plasmas.
When an atom or a molecule is trapped inside any kind of microscopic
cavity, it suffers a spatial confinement that affects its physical and chemical
properties [1,2,4]. The same situation occurs for mesoscopic scale
semiconductor artificial structures like 2-dim. quantum wells [5-7], quantum-
well wires [8,9] and quantum dots [10-14], where impurity or excitonic states
are influenced by the small sizes of these structures. Spatial confinement, 
also called the boxing effect, significantly
influences the bond formation and chemical reactivity inside the cavities
[1,2,4]. There are many natural or artificial cavities that could produce
sensitive effects --- zeolite molecular sieves, fullerenes, or even mono-or
$2$-dimensional cavities formed by large organic molecules [1, 2].

An infinite barrier model is the one most commonly used to study the problems
on spatial confinement. In order to know the energy eigenvalues and
eigenfunctions of such systems, one has to solve the corresponding
Schr\"{o}dinger equation. However, only a few potentials are exactly solvable ;
more so, the number of exactly solvable confined quantum mechanical potentials
is extremely limited. Consequently, one has to apply various approximation
techniques like the direct variational method [3,15], the shifted $1/N$
expansion procedure [12,16] etc. or do the same numerically. It would therefore
be useful to develop formalisms which could study the different interpretations
of the effects of the spatial confinement, in the framework of both solvable
and unsolvable potentials analytically. The recent article by Kr\"{a}hmer {\it
et. al.} [17] gives an outstanding historical account of confined quantum
systems together with the theoretical methods for treating them. It gives
perspective on the different confinement models : hard boxes and soft boxes,
hypervirial theorem approaches etc.

With this goal in mind, we shall develop in this paper, a formalism for 
studying such spatially confined quantum mechanical systems in the 
framework of (usual) WKB (Wentzel-Kramers-Brillouin) and MAF 
(Modified Airy Function) methods. The WKB approximation results in a modified
Bohr-Sommerfeld quantization rule in this context. To test the reliability of 
our formalisms, we shall apply our technique to 3 cases ---

\vspace{0.2cm}

\noindent
(1) \  Confined $1$-dimensional harmonic oscillator (HO), which is one of the
most extensively studied model, both in classical as well as in quantum
mechanics, due to its simplicity and usefulness. It is extremely important for
the quantum mechanical treatment of problems involving vibrations of individual
atoms in molecules and crystals [18]. Even the vibrations of the
electromagnetic field in a cavity can be analysed into harmonic normal modes,
each of which has energy levels of the oscillator type. Moveover, in the study
of $2$-dimensional and $3$-dimensional quantum dots [10], the istropic
parabolic potentials are taken to be of the harmonic oscillator type 

$$\displaystyle{V(\rho) ~=~ \frac{1}{4} ~ \gamma^2 ~ \rho^2 ~~~~{\rm and}~~~~
V(r) ~=~ \frac{1}{4} ~ \gamma^2 ~ r^2 }$$
respectively.

\vspace{0.2cm}

\noindent
(2) \ The $1$-dimensional confined quartic oscillator (QO), which is of
practical interest in molecular physics and in quantum field theory [15,19].

\vspace{0.2cm}

\noindent
(3) \ The interesting problem of a charged particle in an isolated quantum well
structure, subject to an external electric field F [20].

The most widely studied confined systems are the harmonic and anharmonic
oscillators, [2,15,20] and the hydrogen atom [3]. It is worth mentioning here
that Vawter [21] employed the WKB method to study the confined one-dimensional
harmonic oscillator, and obtained pretty accurate eigenvalues. Kr\"{a}hmer
{\it et. al.} [17] investigated the parabolically confined Hydrogen atom in
the WKB approximation and obtained a modified Bohr-Sommerfeld quantization
rule, giving pretty accurate energy eigenvalues. However, we
shall apply our formalism to nonsolvable potentials as well, viz, the boxed-in
quartic oscillator and a charged particle in a box subject to an external
electric field. Recently, a matrix formulation of the Bohr-Sommerfeld
quantisation rule was applied to study bound states in $1$-dimensional quantum
wells [20]. Though our results agree with theirs in case of the $1$-dimensional
harmonic and quartic oscillators, it seems that their analytical result given
for a confined charged particle under the influence of an external electric 
field F, is not compatible with what one would expect for large values of the 
confinement parameter. Very recently, Spehner {\it et. al.} [22] have applied
the WKB method to a similar but more complex quantum system , {\it viz.}
non-interacting electrons constrained to a 2-dimensional domain with
boundaries in the presence of a uniform perpendicular magnetic field.

The organisation of the paper is as follows. In section II we develop a
formalism for the WKB quantisation rule for confined quantum mechanical
systems. In section III we give the MAF formalism for the same. Section IV is
kept for discussions and conclusions.

\section*{II \ WKB formalism for confined systems}

The WKB (Wentzel-Kramers-Brillouin) approximation technique has been extremely
useful in estimating the eigenfunctions and eigenvalues of the Schr\"{o}dinger
equation (yielding exact values for the harmonic oscillator). Our attempt here
is to develop a formalism for studying spatially confined quantum mechanical
systems in the framework of usual WKB quantisation rule, imposing appropriate
boundary conditions. Though this method has been employed by others to solve 
confined  systems, {\it viz.} parabolically confined hydrogen atom [17], 
$1$-d harmonic oscillator [21], and non-interacting electrons in a uniform
magnetic field constrained to a 2-dimensional domain with boundaries [22], 
our approach is different and works for other potentials as well.

We start with the $1$-dimensional Schr\"{o}dinger equation

\begin{equation}
\displaystyle{\{ \frac{d^2}{dx^2} + \Gamma^2 (x) \} ~ \psi(x) ~=~ 0}
\end{equation}
where
\begin{equation}
\displaystyle{\Gamma^2 (x) ~=~ \frac{2m}{\hbar^2} ~ \{ E - V(x) \}} 
\end{equation}

We shall work in units

$$\displaystyle{\hbar ~=~ c ~=~ 2m ~=~ 1}$$

We take the confining potential to be such that it exists only in the region
$-b < x < b$, and is infinite elsewhere. 

\begin{equation}
\left. \begin{array}{lcl}
V(x) &=&\displaystyle{V(x) \quad \mbox{for}\quad - b < x < b}\\ \\
     &&\displaystyle{\infty \quad \mbox{for}\quad | x | > b}
\end{array}\right\}
\end{equation}
This imposes the boundary condition 

\begin{equation}
\displaystyle{\psi (x ~=~ \pm b) ~=~ 0}
\end{equation}

We shall deal with symmetric profiles only, so that 

\begin{equation}
\displaystyle{\Gamma^2 (x) ~=~ \Gamma^2 (-x)}
\end{equation}
and the eigenfunctions are either symmetric or antisymmetric in $x$. Hence

\begin{equation}
\left. \begin{array}{lcl}
\psi(0) &=&\displaystyle{0 \qquad\mbox{for the {\bf antisymmetric function}}}\\
\\ 
\psi^{~\prime} (0) &=&\displaystyle{0 \qquad\mbox{for the {\bf symmetric 
function}}}
\end{array}\right\}
\end{equation}

Let $x_t$ denote the turning point (where $\Gamma^2 (x) = 0$)

\vspace{1cm}

\noindent
{\bf a) \  Turning point inside the box \ ($x_t < b$)}

\vspace{.2cm}

Considering only the half space $0 < x < \infty$, the WKB solution in region I
is 

\begin{equation}
\displaystyle{\psi_I(x) ~=~ \frac{a_1}{\sqrt{\Gamma(x)}} ~sin~ \left(
\int^x_{0} \Gamma(x) dx \right) + \frac{a_2}{\sqrt{\Gamma(x)}} ~cos~
\left(\int^x_{0} \Gamma(x) dx \right)}
\end{equation}

Now $\int^x_{0} \Gamma(x) dx$ can be written as 

\begin{equation}
\displaystyle{\int^x_{0} \Gamma(x) dx ~=~ \int^{x_t}_{0} \Gamma(x) dx -
\int^{x_t}_{x} \Gamma(x) dx ~=~ \alpha - \left(\theta_x + \frac{\pi}{4}
\right)}
\end{equation}

where

\begin{equation}
\displaystyle{\alpha ~=~ \int^{x_t}_{0} \Gamma(x) dx + \frac{\pi}{2}}
\end{equation}

\begin{equation}
\displaystyle{\theta_{x} ~=~ \int^{x_t}_{x} \Gamma(x) dx }
\end{equation}

Then $\psi_I (x)$ can be written as

\begin{equation}
\begin{array}{lcl}
\psi_I(x) &=&\displaystyle{ \frac{1}{\sqrt{\Gamma(x)}}~ (a_1 ~sin~ \alpha + a_2
~ cos ~ \alpha) ~ cos ~ (\theta_x + \pi/4)}\\ \\
&&\displaystyle{+  \frac{1}{\sqrt{\Gamma(x)}}~ (a_2
~sin~ \alpha - a_1 ~ cos ~ \alpha) ~ sin ~ (\theta_x + \pi/4)}
\end{array}
\end{equation}

Making use of the connection formulae for WKB approximation [23]

\begin{equation}
\displaystyle{\frac{2}{\sqrt{\Gamma(x)}} ~ sin ~ \left( \int^{x_t}_x \Gamma(x) dx
+ \pi/4 \right) \equiv \frac{1}{\sqrt{\kappa(x)}} ~exp~ \left( - \int^x_{x_t}
\kappa(x) dx \right)} 
\end{equation}

\begin{equation}
\displaystyle{\frac{1}{\sqrt{\Gamma(x)}} ~ cos ~ \left( \int^{x_t}_x \Gamma(x) dx
+ \pi/4 \right) \equiv \frac{1}{\sqrt{k(x)}} ~exp~ \left( + \int^x_{x_t}
\kappa(x) dx \right)} 
\end{equation}

where

\begin{equation}
\displaystyle{\kappa^2 (x) ~=~ - \Gamma^2 (x)}
\end{equation}

the wave function in region II can be written as 

\begin{equation}
\begin{array}{lcl}
\psi_{II}(x) &=&\displaystyle{(a_1 ~ sin ~ \alpha + a_2 ~ cos ~
\alpha).~\frac{1}{\sqrt{\kappa (x)}} ~ exp ~ \left( \int^x_{x_t}~ \kappa(x) dx
\right)}\\ \\ 
&&\displaystyle{+ (a_2 ~ sin ~ \alpha - a_1 ~ cos ~ \alpha). ~ \frac{1}{2
\sqrt{\kappa (x)}} ~ exp ~ \left( - \int^x_{x_t}~ \kappa(x) dx \right)}
\end{array}
\end{equation}

\vspace{0.2cm}

\noindent
{\bf (i) \ For the antisymmetric function}, the boundary condition (6) gives
$$ \displaystyle{\psi_{I} (0) ~=~ 0}$$

so that from (7),
\begin{equation}
\displaystyle{a_2 ~=~ 0}
\end{equation}

Hence the confinement condition (4) \ (viz. $\psi_{II} (b) = 0$) gives 

\begin{equation}
\displaystyle{e^{\beta} ~sin~ \alpha - \frac{1}{2} ~e^{-\beta} ~ cos ~ \alpha
~=~ 0}
\end{equation}

\vspace{0.2cm}

\noindent
{\bf (ii) \ For the symmetric eigenfunction},
$$ \displaystyle{\psi^{~\prime}_{1} (0) ~=~ 0}$$
so that

\begin{equation}
\displaystyle{a_1 ~=~ 0}
\end{equation}

\newpage

Hence the confinement condition (4) gives 

\begin{equation}
\displaystyle{e^{\beta} ~cos~ \alpha + \frac{1}{2} ~e^{-\beta} ~ sin ~ \alpha
~=~ 0}
\end{equation}

In (17) and (19), $\beta$ stands for 

\begin{equation}
\displaystyle{\beta ~=~ \int^{b}_{x_t} ~ \kappa(x) dx}
\end{equation}

Thus the usual asymptotic WKB quantisation rule gets modified for spatially 
confined $1$-dimensional quantum mechanical systems, and is given 
by eqns. (17) and (19) for the antisymmetric and
symmetric wave functions respectively, provided the turning point is inside the
confining box ; i.e. \ ${x_t} < b$.

\vspace{1cm}

\noindent
{\bf b) \ Turning point outside the box \ ($x_t > b$)}

\vspace{.2cm}

However, for ${x_t} > b$, i.e. in case the turning point lies beyond the
confining length $b$, the boundary condition

\begin{equation}
\displaystyle{\psi_{I} (b) ~=~ 0}
\end{equation}
gives 

\begin{equation}
\displaystyle{\theta_b ~=~ n \pi \qquad\mbox{for the {\bf antisymmetric
function}}\qquad}
\end{equation}
\begin{equation}
\displaystyle{\theta_b ~=~ \left(n + \frac{1}{2} \right) ~ \pi \qquad\mbox{for
the {\bf symmetric function}}\qquad}
\end{equation}

where 
\begin{equation}
\displaystyle{\theta_b ~=~ \int^b_{0} ~ \Gamma (x) dx}
\end{equation}

$$\displaystyle{n ~=~ 0, \ 1, \ 2, \ 3, \ \cdots}$$

We apply our formalism to the $1$-dimensional (i) harmonic and (ii) quartic
oscillators, limited by infinite walls at $x = \pm b$. Such limited oscillator
potentials can be used directly to simulate the lowest excited states of an
oscillating system [15].

\newpage

\noindent
{\bf i) \ Harmonic Oscillator (HO)}

\vspace{1cm}

The $1$-dimensional confined HO is described by the potential

\begin{equation}
\left. \begin{array}{lcl}
V(x) &=&\displaystyle{x^2 \quad \mbox{for}\quad - b < x < b}\\ \\
     &&\displaystyle{\infty \quad \mbox{for}\quad |~ x ~| > b}
\end{array}\right\}
\end{equation}

The turning points are at $x_t = \pm \sqrt{E}$. 

We calculate the energy eigenvalues for various values of the confining
parameter $b$, and observe that our results are far better than those obtained
by the shifted $1/N$ expansion method [16] as shown in Table 1.
It is also observed (from Table 1) that with the increase of the confinement 
parameter the vibrational excitation energy of the oscillator 
decreases, rapidly tending to that of the unlimited oscillator.

\vspace{1cm}

\noindent
{\bf ii) \  Quartic Oscillator (QO)}

\vspace{0.2cm}

The $1$-dimensional confined QO is described by the potential

\begin{equation}
\left. \begin{array}{lcl}
V(x) &=&\displaystyle{x^4 \quad \mbox{for}\quad - b < x < b}\\ \\
     &&\displaystyle{\infty \quad \mbox{for}\quad |~ x ~| > b}
\end{array}\right\}
\end{equation}

The turning points are at $x_t = \pm E^{1/4}$.

The energy eigenvalues are calculated from eqns. (17), (22) and (19), (23) for
the antisymmetric and symmetric functions respectively where eqns. (17), (19) 
are for the turning points lying inside the confining box, and (22), (23) are
for the turning points outside the box. The values are given in Table 2, for 
$ b = 1 $, alongside the values obtained by other methods for comparison, 
explained later on.

\vspace{0.2cm}

\noindent
{\bf iii) \  Infinite quantum well subject to an electric field}

\vspace{0.2cm}

This is the interesting problem of an electric field $F$ being applied on a
particle of charge $- | e |$, bound in an infinite quantum well of width $b$.

\begin{equation}
\left. \begin{array}{lcl}
V(x) &=&\displaystyle{|~e~|~ F x  \quad \mbox{for}\quad 0 < x < b}\\ \\
     &&\displaystyle{\infty \quad \mbox{for}\quad  x > b}
\end{array}\right\}
\end{equation}

The boundary conditions are 

\begin{equation}
\left. \begin{array}{lcl}
\displaystyle{\psi (0)} &=& 0 \\
\displaystyle{\psi (b)} &=& 0 
\end{array}\right\}
\end{equation}
and the turning point is at

\begin{equation}
\displaystyle{x_t ~=~ \frac{E}{|e| F}}
\end{equation}

For $E < | e | F b$, the energy is calculated using eqn. (17), while eqn. (22)
determines the energy for $E > | e | F b$. \\
To simplify calculations, we choose a scaling such that $ |e| ~ F = 1 $ , 
so that 
$$ \left. \begin{array}{lcl}
V(x) &=&\displaystyle{ x  \quad \mbox{for}\quad 0 < x < b}\\ \\
     &&\displaystyle{\infty \quad \mbox{for}\quad  x > b}
\end{array}\right\} $$
The results are quoted in Table 3.

\section*{III \ MAF formalism for confined systems}

Though the WKB quantisation rule works well enough for confined systems, the
WKB solutions are valid in regions far removed from the turning points. In
contrast, the Modified Airy Function (MAF) [23-25] method gives an 
extremely accurate description of both the eigenfunctions as well as 
the eigenvalues in the entire region, including the turning point. 
In this section we shall develop a generalized version of the MAF method, 
which is suitable for spatially confined quantum mechanical systems. 
In this connection it may be noted that Spehner {\it et. al.} [22] have
also considered the Airy function approach to obtain the energies of
non-interacting electrons in a magnetic field, constrained to a 2-dim. domain
with boundaries. However, in our case, the argument in the Airy function 
$ \xi (x) $ is related to $ ( E - V(x) ) ^ {1/2} $ (see equation (31) ) ;
only in case of linear $V(x)$, is $ \xi (x) $ linearly related to $x$.

We start with the $1$-dimensional Schr\"{o}dinger equation (in units $\hbar = c
= 2m = 1$)
$$\displaystyle{\frac{d^2 \psi}{dx^2} + \Gamma^2 (x) \psi ~=~ 0} \eqno {(1)}$$
$$\displaystyle{\qquad\mbox{with}\qquad \Gamma^2 (x) ~=~ E - V (x)} \eqno
{(2)}$$ 

The general MAF solution to (1) is of the form
\begin{equation}
\displaystyle{\psi(x) ~=~ c_1 \frac{Ai {[\xi(x)]}}{\sqrt{\xi^{~\prime} (x)}} + 
c_2 \frac{Bi {[\xi(x)]}}{\sqrt{\xi^{~\prime} (x)}}}
\end{equation}

where $Ai {[\xi(x)]}$, $Bi {[\xi(x)]}$ are the Airy functions defined in
[26] and 
\begin{equation}
\displaystyle{\xi(x) ~=~ - {\left\{ \frac{3}{2} ~ \int_x^{x_t} ~ \Gamma(x) dx
\right\}}^{2/3} ; x < x_t}
\end{equation}

\begin{equation}
\displaystyle{\xi(x) ~=~ {\left\{ \frac{3}{2} ~ \int^x_{x_t} ~ \kappa(x) dx
\right\}}^{2/3} ; x > x_t}
\end{equation}

\begin{equation}
\displaystyle{\Gamma^2 (x) ~=~ - \kappa^2 (x)}
\end{equation}
with $x_t$ being the turning point (i.e. the point where $\Gamma^2(x) = 0$).\\
It is to be noted here that the appearance of the term $ Bi[\xi (x)] $ is
solely due to confinement. It cannot appear in the unconfined case because of
the boundary conditions.

Analogous to the WKB case, we consider only symmetric profiles. Hence the
eigenfunctions are either symmetric or antisymmetric in $x$, obeying boundary
conditions 

\begin{equation}
\left.\begin{array}{lcl}
\psi^{~\prime} (0) &=&\displaystyle{0 \qquad\mbox{for the {\bf symmetric 
function}}}\\ \\
\psi (0) &=&\displaystyle{0 \qquad\mbox{for the {\bf antisymmetric function}}}
\end{array}\right\}
\end{equation}

Also the confining potential is such that it exists only in the region $-b < x
< b$ and is infinite elsewhere. The leads to the boundary condition
\begin{equation}
\displaystyle{\psi (\pm b) ~=~ 0}
\end{equation}

Substitution of (34) and (35) in (30) gives the following eigenvalue equation
(after some straightforward calculations):

\noindent
{\bf i) \ for the antisymmetric function }\\
\begin{equation}
\displaystyle{Ai {[\xi(0)]}. ~ Bi {[\xi(b)]} - Bi {[\xi(0)]}. ~ Ai
{[\xi(b)]} ~=~ 0}
\end{equation}

\noindent
{\bf ii) \ for the symmetric function }\\
\begin{equation}
\displaystyle{\left\{ 4\xi(0). ~ Ai^{~\prime} {[\xi(0)]} + Ai {[\xi(0)]}
\right\}. ~ Bi {[\xi(b)]} - \left\{ 4\xi(0). ~ Bi^{~\prime} {[\xi(0)]} + Bi
{[\xi(0)]} \right\}. ~ Ai {[\xi(b)]} ~=~ 0}
\end{equation}

where $ Ai^ {~\prime} [ \xi (x)], Bi^ {~\prime} [ \xi (x) ]$ are the
derivatives of $Ai[ \xi (x) $ and $Bi [ \xi (x)] $ with respect to $\xi$,
respectively. Thus the eigen energies are obtained from (36) and (37) 
after determining $\xi(0)$ and $\xi(b)$ from (31)-(33) and $Ai (\xi)$, 
$Bi (\xi)$, $Ai^{~\prime} (\xi)$, $Bi^{~\prime} (\xi)$ etc. from the 
formulae given in ref.[26]. We have applied our formalism to all the 
three problems discussed in Section II, and given our results in the 
various Tables. (The symbols are explained later on.)

\section*{IV \ Results and Discussions}

Table 1 gives the first excited state eigenenergies of the 
$1$-dimensional confined HO for different values of the 
confinement parameter $b$. $E(1/N)$ are the energies obtained
by shifted $1/N$ expasion method [16], E(V) are the WKB results of Vawter
[21], E(exV) are the exact numerical energies as given in ref.[21], 
and E(WKB) and E(MAF) are our results by WKB and MAF methods respectively.
E(exact) are the exact numerical ground
state energies of the 3-dimensional HO confined in a spherical box, from
ref. [7]. Strictly speaking, we cannot compare our 1-dimensional results with
the 3-dimensional case. However, since the asymptotic ground state energy of
the 3-dimensional HO ($l=0$ case) coincides with the first excited state 
energy of the 1-dimensional HO, we have assumed the 
comparison to hold even in cases of confinement. In fact, it is easy to observe
from Table 1 that the comparison holds extremely well for $b=1$ and $b=2$, 
{\it i.e.} E(exV) = E(exact). 

\vspace{.5cm}

As is observed from Table 1, both WKB and MAF methods give results far superior
to the shifted $1/N$ expansion procedure, for various values of the confining
parameter $b$. It is interesting to note that our WKB results are better than 
those of the WKB results of Vawter [21] (for $b=2$). It is observed 
that our WKB and MAF results differ very little
from each other. As expected, both WKB and MAF energies tend
rapidly to those of the unconfined oscillator as the well size gets larger. 

\vspace{1cm}

Table 2 gives the results of the confined $1$-dimensional QO, for $b=1$ in 
units of $E_1^{\infty} = \pi ^2 / 8 $. E(BS)
and E(mBS) are the energies by Bohr-Sommerfeld quantisation rule, and a matrix
formulation of the same respectively [20], E(Pwr) and E(Var) are those
obtained by power series expansion [27] and variational methods [15] 
respectively. For the numerical results, E(exact), we quote the 
perturbation results of ref.[28] since it makes no approximation in
deriving the equations, and one can compute the eigenvalues to a high degree
of accuracy by considering a large number of terms of the infinite series.
Our results are denoted by E(WKB) and E(MAF). All the results are quoted after
suitable normalization. The variational method [15] gives the best estimates
of the energies, followed by power series expansion method [27]. As for
comparison among the BS, mBS, WKB and MAF values, the mBS formalism is the
best of the lot. The WKB method (correctly applied) gives energies at least as 
accurate as the BS (Bohr-Sommerfeld) ones, contrary to the claim of Gomes and 
Adhikari [20]. The  MAF method gives marginally better results than 
either BS or WKB approximations.

\vspace{1cm}

Table 3 gives the WKB and MAF energies for a boxed-in charged particle subject
to an external electric field $F$. ($V = | e | F x$  for  $x \leq b$,  and
$\infty$  for  $x > b$). Using a suitable scaling, we have taken 
$ |e| ~ F = 1 $. E(exact) are the accurate energy eigenvalues from ref.[29]. 
For this particular case, the MAF method is supposed to give the exact 
result. However, one has to be very cautious while calculating the values of 
Airy functions and their derivatives, which are highly oscillatory. 
An extremely small error may get magnified in the energy
eigenvalue. As is observed from Table 3, the WKB
quantisation rule gives energies extremely close to the exact values. 
The MAF values are slightly worse than the WKB ones for extremely small 
confinement parameter. As the size of the confining box increases, the MAF
results get better. It is interesting to mention here that 
Gomes and Adhikari [20] attempted this
problem with a matrix formulation of the Bohr-Sommerfeld (mBS) quantisation
rule, claiming WKB does not give good result in this case. Contrary to this
claim we observed that the correct WKB method gives quite accurate values, 
(pretty close to MAF ones). Also the analytical formula given in ref.[20] 
does not give the correct asymptotic behaviour of energy eigenvalues 
for large $b$.(Hence we have not given E(mBS) values for $b > 1$.)
However, the graph presented in their paper apparently gives values 
(obtained by mBS method) quite close to the numerical results.
As $b$ increases, both WKB and MAF energies tend to the 
exact energies for the unconfined case, which is nothing but
the zeroes of the Airy function. 

\vspace{1cm}

To conclude, we have developed a formalism for studying spatially 
confined quantum mechanical systems in the framework of the usual WKB and MAF
methods. Confinement imposes certain boundary conditions which modify the
asymptotic quantization rules. The methods can be applied to 
determine the eigenenergies of non-solvable potentials, bound 
in quantum well structures. To establish the reliability of
our formalism we have calculated the eigenenergies of the confined
$1$-dimensional harmonic and quartic oscillators, and a boxed-in charged
particle in an external electric field, and found our results to be pretty
close to the accurate numerical values, contrary to the claim regarding WKB
method in ref. [20].In fact, our findings encourage us to hope that WKB and
MAF methods would be very useful in studying confined quantum mechanical
systems for various boundary conditions. 

\newpage

\begin{center}
{\bf Table \ 1 \ Harmonic Oscillator ~ $(2m=1)$ }

\vspace{0.2cm}

\begin{tabular}{lllllll}
b   & E($1/N$) & E(WKB) & E(MAF)  & ~E(V)~ &  E(exV)  &  E(exact)\\
0.5 & 40.9612 & 39.5619 & 39.5605 &        &          &  39.5490\\ 
1.0 & 10.5170 & 10.2052 & 10.2050 & 10.20~ &  10.15~  &  10.1510\\
1.5 & ~5.2136 & ~5.1636 & ~5.1635 &        &          &  ~5.0100\\ 
2.0 & ~3.7316 & ~3.5374 & ~3.5368 & ~3.357 &  ~3.529  &  ~3.5296\\ 
3.0 & ~3.0720 & ~3.0129 & ~3.0070 &        &          &  ~3.0122\\ 
5.0 & ~3.0000 & ~3.0000 & ~3.0000 &        &          &  ~3.0000\\ 
\end{tabular}
\end{center}

\vspace{0.3cm}

\begin{center}
{\bf Table \ 2 \ Quartic Oscillator for $b = 1 \ (2m = 1)$}

\vspace{0.2cm}

\begin{tabular}{llllllll}
n   & E(mBS)  & E(BS) & E(Pwr) & E(Var) & E(WKB) & E(MAF) & E(exact)\\
1   & ~2.0901 & ~2.1687 & ~2.0331 & ~2.0314 & ~2.1685 & ~2.1670 & ~2.0317\\
2   & ~8.0901 & ~8.1635 & ~8.0920 & ~8.0855 & ~8.1636 & ~8.1198 & ~8.0860\\
3   & 18.0900 & 18.1628 & 18.0242 & 18.1133 & 18.1628 & 18.1233 & 18.1135\\
4   & 32.0900 & 32.1629 & 32.1313 & 32.1165 & 32.1624 & 32.1612 & 32.1165\\
\end{tabular}
\end{center} 

\vspace{0.3cm}

\begin{center}
{\bf Table \ 3 \ $V = x$ \ for $0 < x \leq b$, and $\infty$ for $x > b \
(2m = 1)$}

\vspace{0.2cm}

\begin{tabular}{lllll}
b   & E(WKB) & E(MAF) & E(mBS) & E(exact)\\
0.3 & 109.8133 & 109.8223 & 109.6461 & 109.8123 \\ 
0.5 & ~39.7286 & ~39.7314 & ~39.4508 & ~39.7283 \\
0.8 & ~15.8222 & ~15.8232 & ~15.3769 & ~15.8208 \\
1.0 & ~10.3717 & ~10.3716 & ~~9.8141 & ~10.3685 \\
1.5 & ~~5.1472 & ~~5.1471 &          & ~~5.1309 \\
2.0 & ~~3.5017 & ~~3.5016 &		 & ~~3.4499 \\ 
3.0 & ~~2.5198 & ~~2.5066 & 		 & ~~2.5090 \\
4.0 & ~~2.3391 & ~~2.3404 &		 & ~~2.3555 \\
5.0 & ~~2.3382 & ~~2.3381 &		 & ~~2.3390 \\
6.0 & ~~2.3382 & ~~2.3381 & 		 & ~~2.3381 \\
\end{tabular}
\end{center} 

\vspace{.5cm}

\section*{Acknowledgment}

The authors are grateful to the referees for some useful comments and
suggestions, without which the paper could not have been witten in the present
form. One of the authors (A.S.) acknowledges financial assistance from the 
Council of Scientific and Industrial Research, India.

\newpage

\section*{References}

\begin{enumerate}
\item[[1]] C. Zicovich-Wilson, W. Jask\'{o}lski and J.H. Planelles. Int. J.
Quantum Chem. {\bf 54} 61 (1995).
\item[[2]] C. Zicovich-Wilson, W. Jask\'{o}lski and J.H. Planelles. Int. J.
Quantum Chem. {\bf 50} 429 (1994).
\item[[3]] S.A. Cruz, E Ley-Koo, J.L. Marin and A. Taylor Armitage. Int. J.
Quantum  Chem. {\bf 54} 3 (1995).
\item[[4]] C. Zicovich-Wilson, A. Corma and P. Viruela. J. Phys. Chem.
{\bf 98} 6504 (1994). 
\item[[5]] S. Chaudhuri. Phys. Rev. B {\bf 28} 4480 (1983).
\item[[6]] M.W. Lin and J.J. Quinn. Phys. Rev. B {\bf 31} 2348 (1985).
\item[[7]] J.L. Marin and S.A. Cruz. Am. J. Phys. {\bf 59} 931 (1991).
\item[[8]] D.M. Larsen and S.Y. Mc Cann.
\item[] a) \ Phys. Rev. B {\bf 45} 3485 (1992)
\item[] b) \ Phys. Rev. B {\bf 46} 3966 (1992)
\item[[9]] J.W. Brown and H.N. Spector.
\item[] a) \ J. Appl. Phys. {\bf 59} 1179 (1986)
\item[] b) \ Phys. Rev. B {\bf 35} 3005 (1987)
\item[[10]] J-L Zhu, J-H Zhao and J-J Xiong. J. Phys. {\it Condens. Matter} {\bf 6}
5097 (1994)
\item[[11]] J-L Zhu and Xi Chen. J. Phys. {\it Condens. Matter}  {\bf 6} L 123 (1994).
\item[[12]] M. El-Said . J. Phys. {\it I} France {\bf 5} 1027 (1995).
\item[[13]] T. Garm - J. Phys. {\it Condens. Matter} {\bf 8} 5725 (1996).
\item[[14]] J-L Zhu, J-Z Yu, Z-Q Li and Y. Kawazoe. J. Phys. {\it Condens. Matter}
{\bf 8} 7857 (1996).
\item[[15]] D. Keeports. Am. J. Phys. {\bf 58} 230 (1990).
\item[[16]] A. Sinha and R. Roychoudhury.  Fizika B {\bf 3} 67 (1994).
\item[[17]] D. S. Kr\"{a}hmer, W. P. Schleich and V. P. Yakovlev.
            J. Phys. A : Math. Gen. {\bf 31} 4493 (1998)
\item[[18]] L.I. Schiff.  Quantum Mechanics (Mc Graw Hill, New York, 1968).
\item[[19]] H.A. Gersch and C.H. Braden. Am. J. Phys. {\bf 50} 53 (1982).
\item[[20]] MAF Gomes and S.K. Adhikari.  J. Phys. B : At. Mol. Opt. Phys. {\bf
30} 5987 (1997).
\item[[21]] R. Vawter. Phys. Rev. {\bf 174} 749 (1968).
\item[[22]] D. Spehner, R. Narevich and E. Akkermans.
            J. Phys. A : Math. Gen. {\bf 31} 6531 (1998).
\item[[23]] A.K. Ghatak, R.L. Gallawa and I.C. Goyal. MAF and WKB Solutions to
the Wave Equations. NIST Monograph 176, Washington, 1991.
\item[[24]] I.C. Goyal, R.L. Gallawa and A.K. Ghatak 
\item[] Opt. Lett. {\bf 16} 30 (1991) 
\item[] Opt. Lett. {\bf 30} 2985 (1991)
\item[[25]] I.C. Goyal, S. Roy and A.K. Ghatak. Can. J. Phys. {\bf 70} 1218
(1992). 
\item[[26]] Abramowitz and I.A. Stegun Hand book of Mathematical Functions,
Dover, 1972.
\item[[27]] R. Barakat and R. Rosner. Phys. Lett. A {\bf 83} 149 (1981)
\item[[28]] R.N. Chaudhuri and B. Mukherjee 
\item[] J. Phys. A : Math. Gen. {\bf 16} 3193 (1983)
\item[] J. Phys. A : Math. Gen. {\bf 17} 3327 (1984)
\item[[29]] V. C. Aguilera-Navarro, H. Iwamoto, E. Ley-Koo and A. H. Zimerman.
            Am. J. Phys. {\bf 49} 648 (1981)
\end{enumerate}
\end{document}